# VirusPKT: A Search Tool For Assimilating Assorted Acquaintance For Viruses

Jayanthi Manicassamy
Department of Computer Science
Pondicherry University
Pondicherry, India.
jmanic2@yahoo.com

P. Dhavachelvan
Department of Computer Science
Pondicherry University
Pondicherry, India.
dhavachelvan@gmail.com

*Abstract*— **Viruses utilize various means to circumvent the immune detection in the biological systems. Several mathematical models have been investigated for the description of viral dynamics in the biological system of human and various other species. One common strategy for evasion and recognition of viruses is, through acquaintance in the systems by means of search engines. In this perspective a search tool have been developed to provide a wider comprehension about the structure and other details on viruses which have been narrated in this paper. This provides an adequate knowledge in evolution and building of viruses, its functions through information extraction from various websites. Apart from this, tool aim to automate the activities associated with it in a self-maintainable, self-sustainable, proactive one which has been evaluated through analysis made and have been discussed in this paper.**

*Keywords – Bioinformatics ;Information Extraction; Search Engin; Viruses; Web-Based.*

## I. INTRODUCTION

Viruses infect cells which elicit response to viral peptides where response plays a critical role in the host's anti-viral immune response [4, 5, 6]. It is necessary to explore the virus related information in terms of structure, evolution, activities in the system for stimulating anti-viral cells response, one have to consider various factors. This involves genetic background of the infected individuals and the viral gene diversifications. These viral genes determine the repertoire presented to the immune system and consequently the immune response [2,3]. Computational efforts have been successful in revealing previously unknown viral. Yet, we hypothesized that many viral have not yet been identified, due various reasons like unawareness, too low sequence similarity, doesn't acquire full knowledge about the viruses [7, 8].

Automatic entity recognition is a hot topic in the mining information is an extensive effort which has already devoted to the identification of viruses and proteins by names. Previous work in this area are only roughly based on only the given search entity for information extraction or Biomedical literature based extraction. There is a lacking in viral based information extraction based on keyword which also includes biomedical literature information. The existing viruses based search tools lacks in information categorization like structure, evolution and functionalities with the immune system.  Apart from this the existing tools and applications lacks behind in extracting information from various viruses related websites.

In this paper in order to explorer virus related information to acquire full knowledge in terms of structure, evolution, various roles and activities a search tool have been developed which extracts information form various websites. This extraction is primarily localized and then made available to the users as per requisition made by the users which have been narrated in section II and III. Results and discussion are been explained in section IV. In section V for the developed tool various quantitative analyses have been carried which have been theoretically explained.

## II. VIRUSPKT ARCHITECTURE

The proposed tool intended to provide the user with multiple facilities within a single tool which is self sustainable with little problem of maintenance capability apart from providing transparency. Figure 1 represents the architecture of the tools functionalities as an overview which has been briefed in this section.

The scope is to have an automated system which fetches data from required websites and customizes it as per the user requirement. Different format data's extracted from various websites are standardized by converting the data's to plain text as per requirement. This involves automatic naming facility for the local copies made, which is considered as one of the major part of other activities involved in processing. The retrieval ability with optimal search results is based on the user queries with a search tag following a specific logic for avoiding result redundancy.

Apart from other facilities Prokut facilitates a socializing link for the communities which have been developed for discussion and knowledge sharing within the communities over specific topics. The tool usage is authorized is specific set of users which is limited with the legal functionality like accessing the databases of the websites whose content is permissible etc.

## III. METHODS

In this section processes carried out but the tool for information extraction from various websites with structures display for acquiring adequate knowledge about the specific virus search made have been narrated. The description has been made phase wise with data flow and transformation along with other required details.

**Figure 1: VirusPKT Architecture**

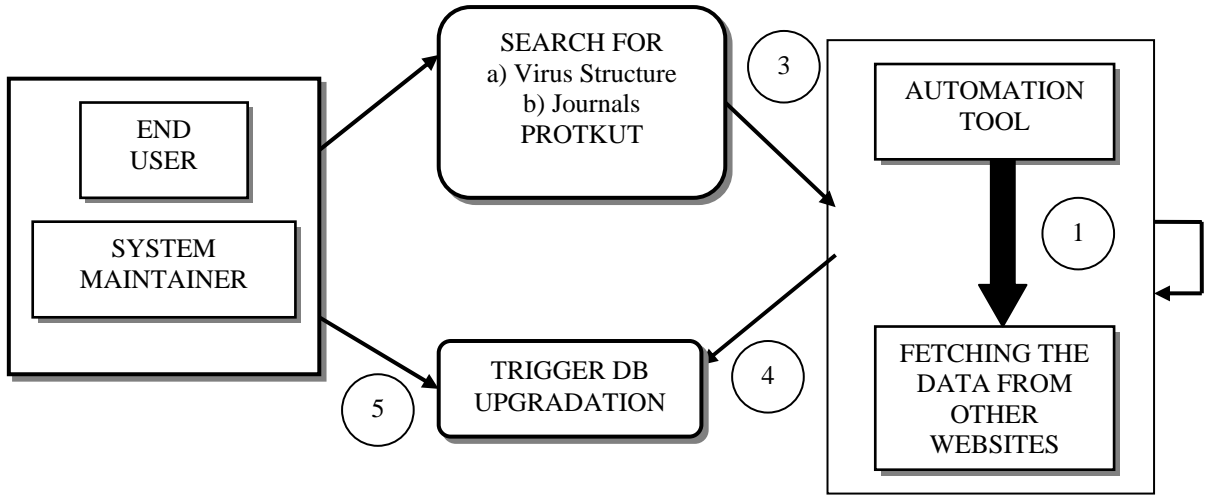

**Figure 2: Data Format Standardize Conversion**

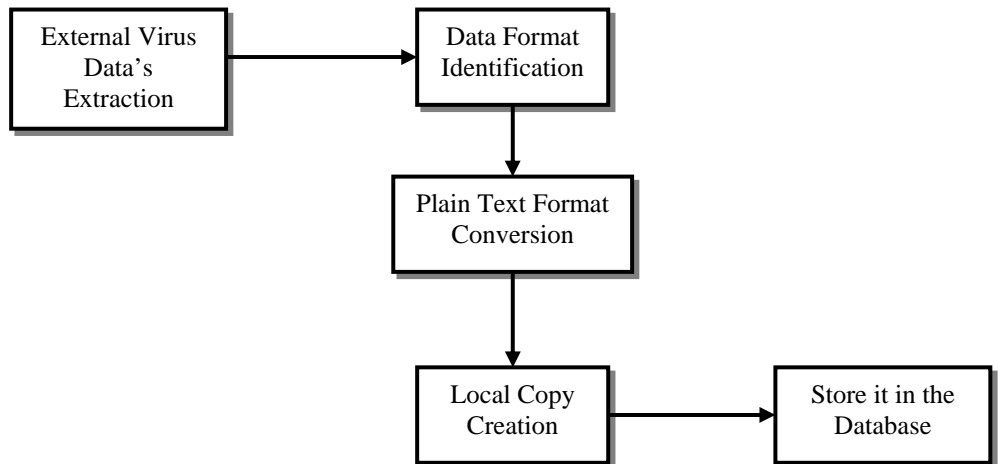

*A. Data Fetching*

This process involves in fetching required data's from required websites with certain specifications for plain text format conversion for fetched data's which has been automated using a PHP script. The fetched data's are converted to plain text format and mad a local copy with is stored in the local database that has been developed for this tool. System Maintainer is responsible for running the specific PHP script for upgrading the database with new information from other required websites.

*B. Format Conversion*

This format conversion process is responsible for converting the data's extracted from various websites with different format like pdf., excel etc. to a standard format which plain text format. Figure 2 represents the overall the dataflow activities involved in data format conversion to a standard one.

**Figure 3: VirusPKT Tool**

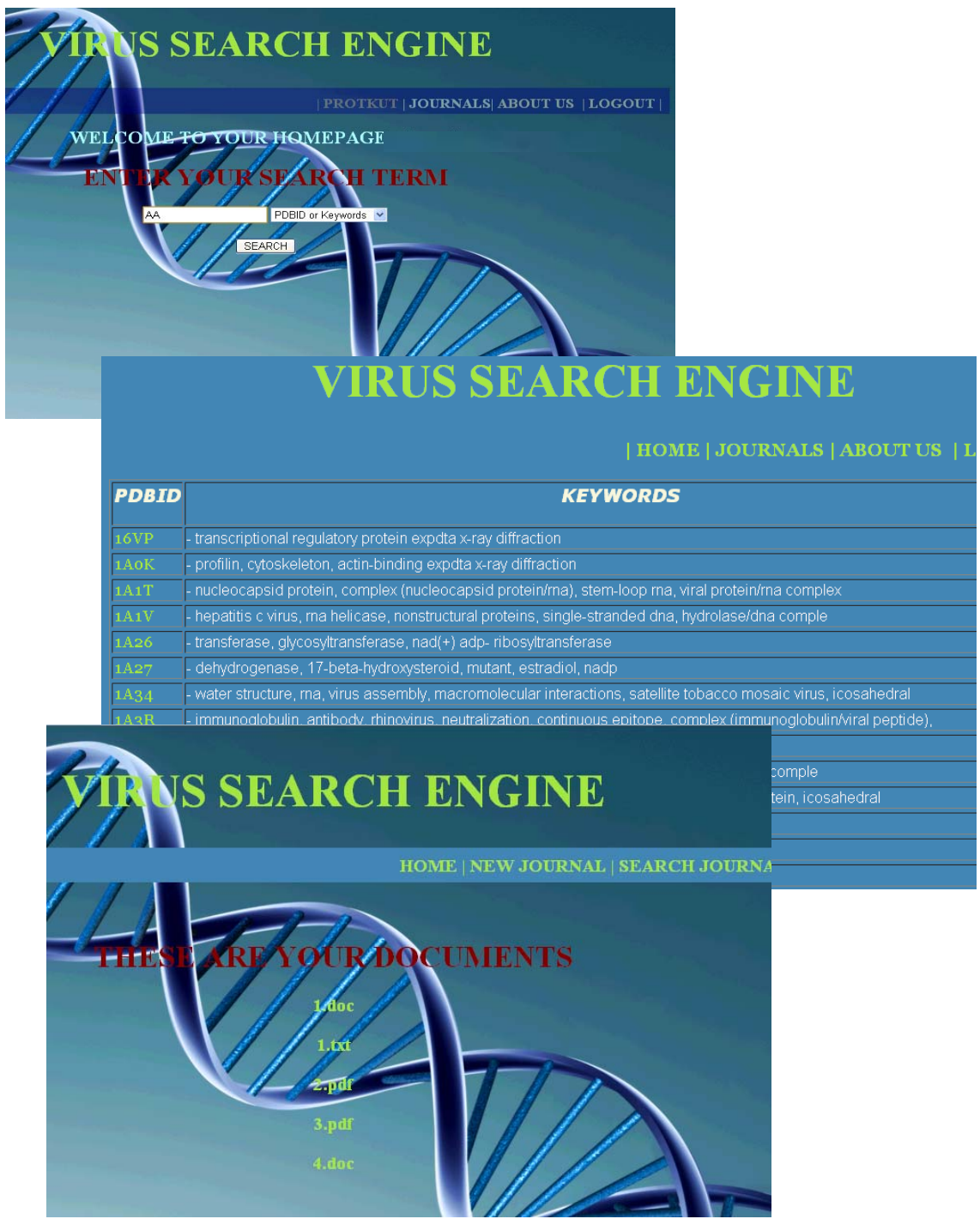

*C. Search*

This is the major task involved by the tool for data's extraction with relates search tags from various websites and databases including journals, articles etc. for the virus related search made based on the user formulated queries, keywords etc. The extraction is made by related and relevant search tags that exist within the journals, articles and websites. The brutal mechanism is to scan the contents of the journals, and if any match found with the search tags the results are retrieved and displayed in the order of their relevancy.

*D. Protkut*

This phase process is an add-on to the tool, which provides functionalities to the community as that of socializing websites. This allows the users in the community to scrap other users in context of discussions regarding the virus details for gaining adequate knowledge. The users are provided with facilities of creating and join communities. Apart from this, facilities provisions for discussions have been also provided for acquiring adequate knowledge about viruses.

IV. USING THE TEMPLATE

The web-based tool provides multiple functionalities which is highly proactive. It does not require much of the technical knowledge to access. The tool extracts information from other websites, via a single PHP script, checks for the format, and converts it to a uniform format. The tool also parses this information into text and image formats. Analyzing the information in several levels enhances the performance of the tool. The user can simply register to access the information. The tool works by using advanced technologies such as AJAX to provide faster and efficient access to the database. An optimal search algorithm is incorporated to search related and relevant results.

Figure 3 gives the VirusPKT tool GUI with outcomes. This tool was developed with a specific cause to make it easily deployable. Session creation is made for authorized users to make the life of the developer easy and highly secure. As many modern browsers use their cache memory for saving time and network bandwidth when the same URL is requested by AJAX repeatedly, there may be a chance that the same response is given by the browser. So, appending a randomly generated number to the URL is incorporated to generate unique URLs every time. For the extracted data's from various databases and websites for the search made data file format is identified and converted to a standard format with is stored locally and made display to the users as per relevancy. Protkut is an add-on with this tool with a slightly different user interface along with scrapbook feature as available in Orkut for communication between users.

The proposed tool is used as an interface to obtain the data from various websites and make it available in a single database. It helps the application to achieve the objectives such as persistent storage, reducing the manual work, faster and easier access to the data, efficient processing of the data in a user-friendly environment. Using this Automation Tool, the users across the world would be able to have a quicker access to the data which establishes an invisible connection loop between various users, which are separated geographically. The virus details, its origin, structure, functions and peculiar behavior are stored in the database and are provided to the user to understand the nature of the virus, so as to tweak its structure to circumvent its harmful effects.

V. VIRUSPKT QUALITATIVE ANALYSIS

For the developed tool various analyses have been made, found to be good which have been theoretically narrated in this section based on performance, security, transparency and maintainability.

*A. Performance*

The web-based tool provided multiple functionalities which is highly proactive uses advanced technologies such as AJAX to provide faster and efficient access to the database. It also uses an optimal search algorithm to make the search results. Apart form this cache memory techniques used in browsers for saving time and network bandwidth when the same URL has also incorporated here. It helps the application to achieve the objectives such as persistent storage, reducing the manual work, faster and easier access to the data's. For the analysis carried out the performance of the tool found to be good.

*B. Security*

Security is a very important criterion to be satisfied by any real-time tools and applications. The developed search tool is also tested exceptionally for its security features, as the data is highly sensitive. Security in prohibiting 'URL Hacking', the session variables are being used, and improper URL redirects the web page to the home page, which demands login to proceed with. Security in handling sessions throughout the application ensures that the session automatically expires after a time gap. Various other security measures are also handled and found to be highly secure for the developed tool

*C. Transparency and Maintainability*

The tool found to be self-sustainable which can be easily upgrade by updating the database, as per the system maintainer's desire which found to easy maintainable. Just by running a PHP script frequently, the tool automatically fetches data from the required websites and journals retrieval from PDB database, thus thereby provides transparency.

VI. CONCLUSION

The main objective of the tool is to provide a wider comprehension about the structure and other detailed note on the virus, which provides adequate knowledge in understanding the build and functions of the viruses. Thus, extracted details from various websites and PDB database by undergoing several processes aiming to automate the activities associated with it in a self-maintainable and self-sustainable in an appreciable way. Additionally, the file is scanned thoroughly using an optimal logic to provide most relevant search results in a minimal amount of time. One of the foreseeable enhancements of the system is to expand the system to connect to other databases and easily retrieve information rather than just the PDB database. But, in order make it a large scale application, it has to be subjected to several types of user-level testing and also the suitability of the application for the same should be affirmed. And the most important aspect is that if the application is secure enough so that it doesn't pose any threat to the related websites.